\documentclass[12pt]{article}
\usepackage{amssymb,amsmath}
\begin{document}

\renewcommand{\thefootnote}{\fnsymbol{footnote}}

\begin{center}
{\large {\bf Determining Static Stresses of Deformed Solitons
 \\[0pt] }}
 \end{center}

\begin{center}
\textbf{V. D. Tsukanov}
\end{center}
\begin{center}{\it Institute of Theoretical Physics, National Science Center}\\[0pt]
{\it ''Kharkov Institute of Physics and Technology''}\\[0pt]
{\it 61108, Kharkov, Ukraine}\\[0pt]
\end{center}

\begin{quotation}
{\small {\rm An equation for the quasi-static soliton ansatz depending on an
arbitrary set of collective variables is covariantly derived on the basis of
the variational approach to the method of collective variables. The field
configuration and the static stresses of the deformed }}$\Phi ^4${\small
{\rm-
 kink that are produced by the excitation of the internal soliton mode are
exactly determined. The kink interaction potential at large distances is
considered for the example of the nonlinear Klein-Gordon system. A general
approach to the problem of exactly determining the intersoliton potential for
the entire set of physically admissible two-soliton configurations is
discussed.}}

\end{quotation}

\renewcommand{\thefootnote}{\arabic{footnote}} \setcounter{footnote}0
\vspace{1cm}

%\begin{center}
\section{Introduction} \label 1
%\end{center}

The dynamic questions in solving some soliton problems are relegated to the
background, at least, at the initial stage, and the problem of determining the
static stresses in the system becomes the key point in the investigation. This
problem directly reduces to finding quasi-static soliton configurations
described using a set of non-Goldstone collective coordinates. In this
context, the problem of determining the intersoliton potential in various
models can be mentioned. The non-Goldstone collective coordinates in these
problems are the separation parameter and also the variables describing the
relative orientation of the soliton pair in the internal space in
multidimensional cases. Furthermore, the non-Goldstone coordinates, which in
essence are nonlinear amplitudes, describe the dynamics of internal soliton
modes. The problem of determining the quasi-static configurations in these
problems has been repeatedly discussed, and the ways to solve it were related
to various model constructions. In this connection, we mention the
investigations of the skyrmion interaction begun as early as 1962
\cite{Sk, JJP, ST, WW, K}. Moreover, the soliton interaction was
also studied for some other models \cite{R2, M, JR}, where some
important properties of intersoliton potentials were established. At, the
same time, it is clear that an approach based on model semiphenomenological
notions does not ensure a comprehensive investigation of the problem, for
which a consistent theoretical scheme is required.

Touching upon this problem, we indicate some difficulties related
to introducing non-Goldstone collective coordinates. For example,
the notion of the distance between extended objects undergoing
mutual deformation acquires its unambiguous visual meaning only if
these objects are at sufficiently large distances from each other.
Therefore, a consistent theory must not only describe the process
of deformation of solitons as they approach each other but also
answer the question of interpreting the ambiguity in determining
the intersoliton distance at the stage where the solitons lose
their identity as a result of this deformation. The difficulties
in constructing a theoretical scheme suitable for describing
intersoliton interaction that are due to the above-mentioned
factor have been extensively mentioned in the literature and were
most distinctly stated in \cite{M2}. Some simple considerations
were later presented in \cite{Ts} in favor of the universal
variational approach to the method of collective variables, which
permits stating various problems of soliton dynamics in the
framework of the Hamiltonian procedure. In \cite{Ts}, the general
equation for quasi-static configurations determined by an
arbitrary set of collective coordinates was also derived. It is
essential that the resulting equation is invariant under
reparameterizing the collective variables, which ensures that the
physical quantities are independent of the way in which the
intersoliton distance is defined.

In Sec. \ref 2, we present a new covariant method for deriving an
equation for the quasi-static ansatz using basic notions related
to Riemannian manifolds. The following two sections are devoted to
elaborating solution methods for this equation in some specific
cases. In Sec. \ref 3, the problem of determining the
configurations and static stresses of the $\Phi ^4$-kink whose
deformation is produced by the excitation of the internal soliton
mode is considered. This problem can be solved exactly. Similar
problems were considered in many papers. The corresponding
references can be found in \cite{BSW, BW}. It is unlikely that
there is a model with a nontrivial potential that admits an exact
solution of the problem of determining the intersoliton
interaction. A solution of this problem that allows calculating an
asymptotic expression for the intersoliton interaction can most
likely be found for an arbitrary model. An example of a
calculation of this kind is given in Sec. \ref 4, where the
equation for the quasi-static ansatz is used to determine the kink
interaction for the nonlinear Klein-Gordon equation in the domain
of large intersoliton distances. We stress that, apart from
independent interest, knowing the intersoliton potential in the
asymptotic domain permits introducing a physically acceptable
gauge in the numerical analysis of the equation throughout the
range of intersoliton distances. In Sec. \ref 5, we discuss the
investigation results and indicate the main elements of the
suggested theoretical scheme.

%\begin{center}
\section{Equation for the quasi-static ansatz} \label 2
%\end{center}

We consider the question of finding the quasi-static ansatz for a
soliton system described by an arbitrary set of collective variables. Let
 \[ L=\sum g_{ik}\dot{q}^i\dot{q}^k-H(q,0) \]
be a nondegenerate Lagrangian quadratic with respect to velocities that
admits soliton solutions, let $q^i$ be the field coordinates, and let
$i\equiv \{i,x\}$  be the set of discrete indices and spatial coordinates.
It is convenient to interpret the basic configuration space as a Riemannian
manifold with a metric $g_{ik}$  generated by the kinetic term. In this case,
if $Q$ are collective variables, then the quasi-static ansatz $q_c^i(Q)$
describing the deformed soliton state can be interpreted as a submanifold
parameterized by the variables $Q$. We assume that this ansatz is determined
by the stationarity condition for the Hamilton function defined on $q_c^i(Q)$
for the system in question relative to arbitrary infinitesimal variations
$\delta q_c^i(Q)$ orthogonal to the manifold surface, i.e.,
\begin{equation}
\frac{\partial H(q_c,0)}{\partial q_c^i}\delta q_c^i=0,
\label{1}
\end{equation}
where the variations $\delta q_c^i(Q)$ satisfy the conditions
\[ g_{ik}(q_c)\frac{\partial q_c^i(Q)}{\partial Q}\delta q_c^k(Q)=0. \]
To cancel the variations in Eq. (\ref{1}), we introduce the operator
${\cal P}$ of projection onto the subspace tangent to the submanifold
$q_c^i(Q)$ at the point $Q$
\begin{equation}
{\cal P}_k^i=g_{kl}(q_c)\frac {\partial q_c^l}{\partial Q}g^{QQ}\frac
{\partial q_c^i}{\partial Q},\qquad {\cal P}^2={\cal P},
\label{2}
\end{equation}
where
\[ g_{QQ}=\frac{\partial q_c^i}{\partial Q}g_{ik}(q_c)\frac{\partial q_c^k}
{\partial Q} \]
is the metric on the submanifold $q_c^i(Q)$ and $g^{QQ}$ is the matrix inverse
to $g_{QQ}$. Because the variations $\delta q_c^i(Q)$ in the orthogonal
directions to $q_c^i(Q)$ are arbitrary, Eq. (\ref{1}) can then be brought to
the form
\[ (1-{\cal P})\frac{\partial H(q_c,0)}{\partial q_c}=0.\]
Substituting expression (\ref{2}) for the operator ${\cal P}$ in this
formula, we obtain
\begin{equation}
\frac{\partial H(q_c,0)}{\partial q_c^i}=g_{ik}(q_c)\frac{\partial q_c^k}
{\partial Q}g^{QQ}\frac{\partial E(Q)}{\partial Q},
\label{3}
\end{equation}
where $E(Q)\equiv{H(q_c(Q),0)}$ is the energy of the quasi-static
configuration. By the definition of the tensor $g^{QQ}$, the right-hand side
of Eq. (\ref{3}), which takes the existence of static stresses in the
collective subsystem into account, is invariant under an arbitrary
nondegenerate change of variables $Q\rightarrow Q^{\prime }(Q)$. This
indicates that the variational procedure under consideration is covariant. In
this connection, we note that in relation to the problem of determining
two-soliton configurations, for example, Eq. (\ref{3}) exactly describes
these configurations in the domain of distances where the deformed solitons
lose their identity. The choice of the gauge only determines the specific
form of the parameterization describing the configurations. For convenience,
it can be required that the separation parameter coincide with the natural
intersoliton distance if this distance notably exceeds the soliton size. We
note that if the collective coordinates are Goldstone variables, i.e., if the
coordinates $Q$ coincide with the degeneration parameters of vacuum
solutions, then the energy $E$ does not depend on $Q$, and Eq. (\ref{3})
becomes the equation ${\partial H(q_c,0)}/{\partial q_c^i}=0$ for static
solutions.

%\begin{center}
\section{Static stresses of the deformed $\Phi ^4$-kink.} \label 3
%\end{center}

We use Eq. (\ref{3}) to determine the configuration of the deformed kink in
the $\Phi ^4$ model. The Lagrangian of the model and the corresponding static
kink solution $u_c(x-X)$ defined to within the spatial coordinate $X$ have
the forms
\begin{equation}
 L=(1/2)\int dx\left( \dot{\Phi}^2(x)-\Phi
^{\prime 2}(x)-(\Phi ^2(x)-1)^2\right) ,\qquad u_c(x)=\tanh x.
\label{4}
\end{equation}
The eigenfunctions of the operator ${\cal L}$ describing linear fluctuations
on the background of the kink solution include the mode $ \sigma (x)$
belonging to the discrete-spectrum states,
\begin{equation}
{\cal L}\sigma (x)=3\sigma (x),\quad \sigma (x)=\sqrt{3/2}\frac{\tanh
x}{\cosh x}, \quad {\cal L}=-\frac{\partial^2}{\partial x^2}+4-6\cosh^{-2}x.
\label{5}
\end{equation}
In describing nonstationary states, it is usually assumed that the
$\Phi ^4$-kink possesses an internal degree of freedom whose small
oscillations can be identified with the mode $\sigma (x)$. Therefore, apart
from the translational coordinate $X$, the set of collective coordinates
describing the kink state must also include the variable $\tau$ related to the
excitation of the internal mode of the $\Phi ^4$-kink. For this set of
collective coordinates, we can expect that the directing vectors
$\Phi _c^{\prime }\equiv \partial \Phi _c/\partial x$ and
$\Phi _{c\tau }\equiv \partial \Phi _c/\partial \tau $  constructed on the
solutions of Eq. (\ref{3}) are orthogonal in the tangent space,
\begin{equation}
\int dx\Phi _c^{\prime }(x)\Phi _{c\tau }(x)\equiv <\Phi _c^{\prime }\Phi
_{c\tau }>=0.
\label{6}
\end{equation}
Also, taking into account that the energy is independent of the cyclic
variable, we can represent Eq. (\ref{3}) in the case under consideration as
\begin{equation}
-\Phi _c^{\prime \prime }(x)+2\Phi _c(x)(\Phi _c^2(x)-1)=\frac{\partial\Phi
_c(x)}{\partial\tau}<\Phi _{c\tau }^2>^{-1}\frac{\partial E}{\partial \tau} .
\label{7}
\end{equation}
The right-hand side of Eq. (\ref{7}) is invariant under the
reparameterization $\tau \rightarrow \tau ^{\prime }(\tau )$. If the
collective variable is fixed using the gauge \begin{equation}
 -<\Phi _{c\tau }^2>^{-1}\partial E/\partial\tau =1,
\label{8}
\end{equation}
then Eq. (\ref{7}) becomes the nonlinear diffusion equation
\begin{equation}
\frac{\partial\Phi
_c(x)}{\partial\tau}-\Phi _c^{\prime \prime }(x)+2\Phi_c(x)(\Phi _c^2(x)-1)=0.
\label{9}
\end{equation}
Equation (\ref{9}) was investigated in \cite{NW}, where its exact solutions
were found. In particular, the solutions of the type of a deformed kink that
tend to $+1$ as $x\rightarrow +\infty $ can be written in the form
\begin{equation}
\Phi _c(x,\tau )=\frac{e^{2x}-1}{e^{2x}+1\pm e^{x-3\tau }}.
\label{10}
\end{equation}
Integrating, we can easily verify that these solutions satisfy orthogonality
condition (\ref{6}) and gauge condition (\ref{8}). We now note that the
physical domain of small deformations, where solutions (\ref{10}) only
slightly differ from static kink solution (\ref{4}), corresponds to $\tau
\rightarrow \infty $. In other words, the collective variable $\tau $ defined
by gauge condition (\ref{8}) is inconvenient from the physical standpoint.
It plays only an intermediate role as it permits representing Eq. (\ref{7})
in the form of a nonlinear diffusion equation. It is therefore advisable to
combine the two solutions in (\ref{10}) to form one solution,
\begin{equation}
\Phi _c(x,q)=\frac{e^{2x}-1}{e^{2x}+1-\sqrt{6}qe^x}\thickapprox
 u_c(x)+q\sigma (x)+O(q^2),
\label{11}
\end{equation}
where the new collective variable $q$, which takes both positive and negative
values, has the meaning of the amplitude of the internal mode under small
deformations \cite{Ts}. The configuration energy $E(q)\equiv H(\Phi _c,0)$
determining the static stresses of the deformed kink has the form
\[
E(q)=\frac 43+\frac{\overline{q}^2}{1-\overline{q}^2}-\frac{\overline{q}^3}{
(1-\overline{q}^2)^{3/2}}\arccos \overline{q}\thickapprox \frac 43+\frac
32q^2+O(q^3),\qquad \overline{q}=\sqrt{3/2}q.
\]
Here, the first term is the kink mass, and the quadratic term with respect to
$q$ describes the potential of a harmonic oscillator with the internal mode
frequency $\omega^2=3$. The range of the collective variable where the energy
remains real is confined to the interval $-\sqrt{2/3}<q<+\sqrt{2/3}$. Solution
(\ref{11}) has no singularities with respect to the spatial coordinate in
this range of $q$. We also stress that the choice of the physical collective
variable tending to the amplitude of the internal mode in the limit of small
deformations is ambiguous. Namely, it is determined for convenience and is
connected with explicit form (\ref{10}) of the solution for the deformed
kink in the situation in question.

%\begin{center}
\section{Kink interaction.} \label 4
%\end{center}

An important role in studying the properties of soliton states is played by
the solutions of Eq. (\ref{9}) that are defined by the asymptotics
$\Phi _c(x)\rightarrow -1$ as $x\rightarrow \pm\infty $. These solutions
describe the kink-antikink configurations and permit establishing the
properties of the intersoliton interaction. In this case, the collective
variable relates to the separation parameter $r$, $2r$ is the intersoliton
distance. The excitations of internal modes are not taken into account.
Unfortunately, it seems that exact solutions of Eq. (\ref{9}) with given
boundary conditions have not yet been found \cite{KT}. Approximate solutions
in the domain of large distances can be obtained in the general form if the
corresponding equation
\begin{equation}
-\Phi _c^{\prime \prime }(x)+U^{\prime }(\Phi _c(x))=\frac{\partial\Phi
_c(x)}{\partial r}<\Phi _{cr }^2>^{-1}\frac{\partial E}{\partial r},
\label{12}
\end{equation}
for the nonlinear Klein-Gordon model is considered instead of (\ref{7}).
Here, $U(\Phi )$ is the potential of the model. If the origin is placed at
the center of the bisoliton pair and its symmetry $\Phi _c(x,r)=\pm
\Phi_c(-x,r)$ is taken into account, then the entire configuration can be
described by considering the domain $x\leqslant 0$. The even functions
describe the kink-antikink states and the odd ones describe the kink-kink
states (in the case of a periodic potential $U(\Phi )$). If the intersoliton
distance is large, then to the left of the point $x=0$, the ansatz
$\Phi_c(x,r)$ only slightly differs from the one-kink solution $u_c(x+r)$ of
the static Klein-Gordon equation
\[ -u_c^{\prime \prime}(x)+U^{\prime}(u_c(x))=0. \]
Therefore, representing the ansatz $\Phi _c(x,r)$ in the form
\begin{equation}
\Phi _c(x,r)=u_c(x+r)+\eta (x+r,r),\qquad x\leqslant 0,
\label{13}
\end{equation}
where $\eta (x+r,r)$ tends to zero as $r\rightarrow \infty$, and taking into
consideration that in the domain of large values of $r$,
\begin{equation}
<\Phi _{cr}^{2}>\approx2\int\limits_{-\infty }^\infty dxu_c^{\prime
2}(x)\equiv2M,
\label{14}
\end{equation}
where $M$ is the kink mass, we derive an equation for $\eta (x)$ from
(\ref{12}),
\begin{equation}
{\cal L}\eta (x)=\frac1{2M}\frac{\partial E}{\partial r}u_c^{\prime
}(x),\qquad x\leqslant r,\qquad {\cal L}\equiv -\frac{\partial ^2}{\partial
x^2}+U^{\prime \prime }(u_c(x)).
\label{15}
\end{equation}
Here, $\partial E/\partial r$ is a small parameter in the domain of large
values of $r$. Multiplying Eq. (\ref{15}) by the zero mode $u_c^{\prime }(x)$
of the operator ${\cal L}$ and integrating by parts from $-\infty$ to $x$ ,
we can reduce the order of the equation. As a result, we obtain the
linear first-order equation for $\eta (x)$
\begin{equation}
-u_c^{\prime }(x)\eta ^{\prime }(x)+u_c^{\prime \prime }(x)\eta (x)=\frac{
M(x)}{2M}\frac{\partial E}{\partial r},\quad
M(x)=\int\limits_{-\infty }^xdx^{\prime }u_c^{\prime 2}(x^{\prime }).
\label{16}
\end{equation}
The solution of Eq. (\ref{16}) can be written as
\begin{equation}
\eta (x)=u_c^{\prime}(x)\left ( \frac{\eta (r)}{u_c^{\prime
}(r)}+\frac 1{2M}\frac{\partial E}{\partial\,
r}\int\limits_x^{r}dx^{\prime}\frac {M(x^{\prime })}{
u_c^{\prime2}(x^{\prime })}\right ) . \label{17}
\end{equation}
The function $\eta (r)$ playing the role of the integration
constant in (\ref{17}) is found from the boundary conditions. For
the kink-antikink and kink-kink pairs, these conditions have the
respective forms $\Phi_c^{\prime }(0,r)=0$ and $\Phi _c(0, r)=0$.
Using these conditions and Eq. (\ref{16}), we obtain the following
expressions for the boundary values $\eta (r)$ and $\eta ^{\prime
}(r)$ in these two cases:
\begin{equation}
\eta ^{\prime }(r)=-u_c^{\prime}(r),\qquad \quad \eta (r)=\frac
1{u_c^{\prime \prime}(r)}\left(\frac{M(r)}{2M} \frac {
\partial E}{\partial\, r}-u_c^{\prime2}(r)\right)
\label{18}
\end{equation}
for the kink-antikink pair and
\begin{equation}
\eta (r)=-u_c(r),\quad \eta ^{\prime }(r)=-\frac
1{u_c^{\prime}(r)}\left( \frac {M(r)}{2M} \frac{\partial
E}{\partial\, r}+u_c^{\prime \prime }(r)u_c(r)\right) \label{19}
\end{equation}
for the kink-kink pair. Solution (\ref{17}) describes the distortion of the
kink under the effect of the second soliton when the distance between the
kinks substantially exceeds their sizes and depends on the small parameter
$\partial E/\partial r$, which determines the interaction of the solitons. We
note that formulas (\ref{17}-\ref{19}) include terms of higher order than the
leading asymptotic terms with respect to $r$. However, because the function
$\eta (x)$ contains competing spatial exponentials, it is inadvisable to
select these terms in considering the field variable $\eta (x)$. Formulas
(\ref{17}-\ref{19}) should therefore be used to correctly determine the
leading approximation term for the integrated characteristic, the interkink
potential $E(r)$. Because $(1/2)u_c^{\prime 2}=U(u_c)$, substituting
(\ref{13}) in the formula for the energy $E(r)=H(\Phi _c,0)$, expanding with
respect to $\eta (x)$, and integrating by parts with regard to Eq. (\ref{16})
result in
\begin{multline}
E(r)-2M=-2\int\limits_{r}^\infty dxu_c^{\prime 2}(x)
+\frac 1{2M}\frac{\partial E}{ \partial r}\int\limits_{-\infty }^{r}dx\eta
(x)u_c^{\prime }(x)\\
+\eta (r)(2u_c^{\prime }(r)+\eta ^{\prime }(r)).
\label{20}
\end{multline}
Substituting expression (\ref{17}) for $\eta (x)$ in (\ref{20}) and using
boundary values (\ref{18}) and (\ref{19}) for the kink-antikink and
kink-kink pairs, we see that the linear terms with respect to $\partial
E/\partial r$ cancel. In this case, relation (\ref{20}) becomes
\begin{equation}
E(r)-2M=A(r)+B(r)\left( \frac 1{2M}\frac{\partial E}{\partial r}\right) ^2,
\label{21}
\end{equation}
where the functions $A(r)$ and $B(r)$ are given by the formulas
\begin{eqnarray*}
A(r) &=&-\frac{u_c^{\prime 3}(r)}{u_c^{\prime \prime }(r)}-2\int\limits_
{r}^\infty dxu_c^{\prime 2}(x), \\ B(r) &=&\frac{M^2(r)}{u_c^{\prime}(r)
u_c^{\prime \prime}(r)}+\int\limits_{-\infty }^{r}dx\frac{M^2(x)}
{u_c^{\prime 2}(x)}
\end{eqnarray*}
for the kink-antikink pair and
\begin{eqnarray*}
A(r) &=&u_c^2(r)\frac{u_c^{\prime \prime}(r)}{u_c^{\prime }(r)}+2\int\limits_
{r}^\infty dxu_c(x)u_c^{\prime \prime}(x), \\ B(r) &=&\int\limits_{-\infty }^
{r}dx\frac{M^2(x)}{u_c^{\prime 2}(x)}
\end{eqnarray*}
for the kink-kink pair. The asymptotic expression for the kink solution has
the form
\[ u_c(x)\vert_{x\rightarrow \pm \infty }{\rightarrow }u_{\pm }\mp{a\exp (\mp
mx)}, \]
where $m=\sqrt{U^{\prime \prime }(u_{\pm })}$ is the pion mass. In a case of
an antisymmetric kink-kink configuration the vacuum constant $u_{+}$ is equal
to zero. Taking into consideration that $M(\infty )=M$ and $M(-\infty )=0$
and calculating the leading approximation terms with respect to $e^{-mr}$ for
the coefficients $A(r)$ and $B(r)$, we see that the functions $A(r)$ vanish
and $B(r)$ coincide to within the sign. Therefore the leading approximation
for Eq. (\ref{21}) has the form
\[ E(r)-2M=\mp \frac 1{8m^3a^2}e^{2mr}\left( \frac{\partial
E}{\partial r} \right) ^2. \]
whence we find the interaction energy of the
bisoliton pair at large distances,
\begin{equation}
E(R)=2M\mp 2ma^2e^{-2mr}.
\label{22}
\end{equation}
This formula describes attraction (repulsion) for the kink-antikink
(kink-kink) configuration. The expression for the energy in (\ref{22}) in the
case of the $\Phi ^4$ and sine-Gordon systems was found in \cite{R2} using an
approximate equation of type (\ref{12}) whose right-hand side was imitated
with a phenomenological delta-shaped source \cite{R}.

To determine the soliton configuration and interaction throughout the range of
distances, the gauge should be fixed, and Eq. (\ref{12}) should be solved
numerically. We note that the representation of the solution in form
(\ref{13}) partly fixes the gauge in the domain of large intersoliton
distances. Formulas (\ref{14}) and (\ref{22}) obtained using expansion
(\ref{13}) can be used to introduce the natural gauge
\begin{equation}
<\Phi _{cr}^2>^{-1}\frac{\partial E}{\partial r}=\pm
\frac{2{m^2}{a^2}}{M}e^{-2mr},
\label{23}
\end{equation}
ensuring the coincidence of the doubled parameter $r$ with the true
intersoliton distance in the asymptotic domain $r\rightarrow \infty $. After
gauge (\ref{23}) is fixed, Eq. (\ref{12}) takes the form of a nonlinear
diffusion equation,
\begin{equation} -\Phi _c^{\prime \prime }(x)+U^{\prime}(\Phi _c(x))=
\pm \frac{2m^2a^2 }{M}e^{-2mr}\frac{\partial \Phi _c(x)}{\partial r}.
\label{24}
\end{equation}
Equation (\ref{24}) can be used to find the corrections to potential
(\ref{22}) in the asymptotic domain for the given gauge, and its numerical
solution permits reproducing the entire set of physically admissible
two-soliton configurations. We stress that solving this equation in the case
of the kink-antikink configuration corresponds to describing diffusion in the
direction of positive "time" $-r$. This process is stable, and the mutual
approach in the kink-antikink pair terminates with its "annihilation" under
which $ \Phi _c(x)$ goes into the vacuum constant \cite{KT}. Actually, the
calculation of the evolution of the kink-antikink pair performed in \cite{KT}
in relation to the investigation of the nonlinear diffusion equation is an
example of rigorous determination of the entire set of two-soliton
configurations in problems related to studying intersoliton interaction on
the basis of Eq. (\ref{3}).

Constructing the kink-kink configuration reduces to studying diffusion in the
direction of negative "time" $r$. In the domain of large intersoliton
distances, the presence of nonlinearity in Eq. (\ref{24}) stabilizes the
instability characteristic of this procedure in the case of the linear
theory. The stabilization weakens during the mutual approach, and the
limiting field configuration is determined by the point where the stability
of the process terminates.

%\begin{center}
\section{Conclusion} \label 5
%\end{center}

We have covariantly derived an equation for a quasi-static ansatz on the basis
of the variational approach. A characteristic feature of this approach is the
invariance under reparameterization with a change of collective variables. The
cyclic variables or the variables related to the degeneration of vacuum states
are introduced in the standard way. The arbitrariness in the choice of the
non-Goldstone variables can be eliminated based on some physical
considerations in the domain where these variables can be given an immediate
physical interpretation. Otherwise they are chosen for convenience.

As an example, we have considered the static deformations of the $\Phi ^4$-kink
related to excitation of the internal mode. This problem admits an exact
solution. The configurations of the deformed kink, its energy, and the
physically admissible range of the amplitude of nonlinear oscillations have
been determined. Determining the intersoliton potential for the nonlinear
Klein-Gordon equation has also been touched upon. Linearizing the equation for
the quasi-static ansatz in the domain of large distances permits establishing
the deformation of the kinks and the asymptotic character of their interaction,
which allows introducing the natural gauge and representing the equation for
the quasi-static ansatz as a nonlinear diffusion equation suitable for
numerical determination of the entire set of physically admissible two-soliton
configurations

In relation to determining the intersoliton interaction, we stress that
although two-soliton configurations have been comprehensively studied using
various qualitative approaches for the majority of actual models,
investigations on the basis Eq. (\ref{3}) permit including the exact effect
of soliton deformation at small distances, i.e., in the domain most sensitive
to the defects characteristic of artificial field constructions. This is
particularly important, for instance, in the Skyrme model because the
nucleon-nucleon potential in this domain has an insignificant minimum forming
the nuclear coupling energy. Although Eq.(\ref{3}) does not take the spin
effects into account, the study of skyrmion-skyrmion interaction on the basis
of this equation is a useful stage in the procedure of determining the
nucleon-nucleon potential.
\newline
\indent The author is grateful to P. 0. Mchedlov-Petrosyan and V.V. Gann for
a number helpful discussions.

\vspace{0,5 cm}

\end{document}